\documentclass[12pt]{article}
\usepackage{amsmath}
\def\ie{\hbox{\it i.e.}}        
\def\eg{\hbox{\it e.g.}}        
\def\etal{\hbox{\it et al.}}

\let\vev\VEV

\newlength{\myem}
\newcounter{mysubequation}[equation]

\def\lesssim{\mathrel{\mathpalette\vereq<}}

\makeatletter
\def\vereq#1#2{\lower3pt\vbox{\baselineskip1.5pt \lineskip1.5pt
\ialign{$\m@th#1\hfill##\hfil$\crcr#2\crcr\sim\crcr}}}
\makeatother

\newcommand{\be}{\begin{equation}}
\newcommand{\ee}{\end{equation}}
\newcommand{\ep}{$\epsilon$}
\newcommand{\epp}{$\epsilon '$}
\newcommand{\mep}{\epsilon}
\newcommand{\mepp}{\epsilon '}

\renewcommand{\thefootnote}{\fnsymbol{footnote}}
\title{
\normalsize
\begin{tabular}[t]{l}\end{tabular}
\hspace*{\fill}
\begin{tabular}[t]{l}
SNS--PH/99--1 \\
OUTP--99--19P \\
hep-ph/9903460 \\
March 1999
\end{tabular}
\vspace{3\baselineskip}\\\LARGE\bfseries 
Neutrino mixings from a U(2) flavour symmetry\thanks{This work was
supported in part by the TMR Network under the EEC Contract No.\
ERBFMRX--CT960090.}
\smallskip} 
\author{\begin{minipage}[t]{0.8\textwidth}
\large\centering 
\textbf{\smallskip Riccardo Barbieri$^{\, a}$, Paolo
Creminelli$^{\, a}$, \\ Andrea Romanino$^{\, b}$}%
\medskip\\
\normalsize 
{\em $\mbox{}^a$ Scuola Normale Superiore and INFN, Sezione di Pisa, \\
I--56126 Pisa, Italy}
\\[0.1\baselineskip] 
{\em $\mbox{}^b$ Department of Physics, Theoretical Physics,
University of Oxford, Oxford OX1 3NP, UK}
\end{minipage}}
\date{}

\begin{document}

\bigskip
\maketitle
\begin{abstract}\normalsize\noindent
We extend a previously developed description of the flavour parameters
in the charged fermion sector, based on a U(2) flavour symmetry, to
include two main features of the neutrino sector seemingly implied by
recent data: a large mixing angle $\theta_{\mu\tau}$ and a large
hierarchy in the neutrino squared mass differences.  A unified
description of quark and lepton masses and mixings emerges. The
neatest quantitative predictions are for elements of the unitary
mixing matrix in the lepton sector:
\[
{\displaystyle \left|V_{\mu 1}^\ell\right| = \left|\frac{V_{e3}^{\ell}}{V_{\mu3}^{\ell}}
\right| = \left|\frac{V_{e2}^{\ell}}{V_{\tau 3}^{\ell}}
\right| = 
\sqrt{\frac{m_e}{m_{\mu}}}},
\]
which go together with the analogous relations in the quark sector:
\[
{\displaystyle 
\left|\frac{V_{ub}^{q}}{V_{cb}^{q}}\right| = \sqrt{\frac{m_u}{m_{c_{\hspace{1pt}}}}}}, \quad 
{\displaystyle \left|\frac{V_{td}^{q}}{V_{ts}^{q}}\right| = \sqrt{\frac{m_d}{m_{s_{\hspace{1pt}}}}}}.
\]
\end{abstract}
\normalsize\vspace{\baselineskip}
\clearpage

\renewcommand{\thepage}{\arabic{page}}
\setcounter{page}{1}
\renewcommand{\thefootnote}{\arabic{footnote}}
\setcounter{footnote}{0}

\section{Introduction}
If neutrinos indeed oscillate, as seemingly implied, at different
level of evidence, by several experiments, the number of flavour
parameters in the current description of particle physics increases
from 13 to 22 in the case of three light Majorana neutrinos: 3
neutrino masses, $m_i$, and a unitary mixing matrix in the charged
leptonic weak current with 3 angles and 3 physical phases. Is there an
overall rationale behind these parameters within the current paradigm
of particle physics?

Unlike the case of the quark sector, where most of the parameters are
known, sometimes even with significant precision, the available
information in the neutrino sector is still very scanty. The
interpretation of the atmospheric neutrino anomaly in terms of
neutrino oscillations requires a large mixing angle $\theta_{23}$
between $(\mu, \tau)$ and two of the neutrino mass eigenstates
$\nu_2$, $\nu_3$~\cite{SK}. Furthermore the squared mass difference
between these two states, $\Delta m^2_{23}$, is in the $10^{-2} \div
10^{-3}\: {\rm eV^2}$ range and, likely, significantly larger than the
other independent neutrino mass squared difference, $\Delta m^2_{23}
\gg \Delta m^2_{12}$, as suggested by solar neutrino
experiments~\cite{sun}. Not much else is reliably known at present,
except that no simultaneous explanation is possible, in a three
neutrino oscillation picture, of the LSND result~\cite{LSND} together
with the atmospheric and solar neutrino anomalies, even if the energy
dependence of the suppression of the solar neutrino flux is
neglected~\cite{BHSSW}.

In spite of this scanty information several attempts have been made to
explain the largeness of $\theta_{23}$ in terms of flavour
symmetries. This becomes non trivial if one wants a parametric rather
than accidental explanation of $\theta_{23} ={\cal{O}}(1)$ and, at the
same time, of the hierarchy $\Delta m^2_{23} \gg \Delta m^2_{12}$. It
is nevertheless possible, using abelian or non abelian symmetries, in
a number of different ways. A next natural step, in the same general
direction, is to search for a coherent and testable overall
description of quark and lepton masses and mixings, which includes
these features of neutrino physics.

A prevailing attitude taken in the literature~\cite{SU5,textures} with
respect to this problem can be phrased in SU(5) language as
follows. Denoting the three families of SU(5) 10-plets and 5-plets by
$T_i$, $\bar{F}_i$, $i$=1,2,3, respectively, let us assume that the
dominant effective Yukawa couplings are:
\begin{equation}
\label{YukawaSU5}
\lambda^T_{ij} T_iT_jH, \qquad  \lambda^F_{ij} T_i\bar{F}_j\bar{H}, \qquad
\frac{\lambda^N_{ij}}{M}\bar{F}_iHH\bar{F}_j
\end{equation} 
where $H$, $\bar{H}$ are the usual Higgs 5-plets and $M$ is a heavy scale,
possibly the Planck scale. To the extent that this is true, in the basis for
$\bar{F}$ and $T$ where $\lambda^T$ and $\lambda^N$ are diagonal, writing 
$\lambda^F$ in terms of its diagonal form as:
\begin{equation}
\label{lambda5}
\lambda^F=\left(V^q\right)^t\lambda^F_{\rm diag}V^{\ell},
\end{equation}
the unitary matrices $V^q$ and $V^{\ell}$ represent the mixing matrices in the 
quark and lepton weak charged currents respectively. Suppose now that a 
suitable family symmetry, \eg\ an abelian U(1), gives $\lambda^F$ in the 2,3
sector of the form:
\begin{equation}
\label{asymmetry}
\frac{\lambda^{F (2,3)}}{\lambda^F_{33}} = \left(\begin{array}{cc} {\cal {O}}
(\epsilon) & {\cal {O}}(\epsilon) \\ {\cal {O}}(1) & 1 \end{array} \right),
\end{equation}  
where $\epsilon$ is a small symmetry breaking parameter of order $m_s/m_b$ or 
$m_{\mu}/m_{\tau}$. It is evident from (\ref{asymmetry}) that the right 
diagonalization matrix gives a large mixing angle in the lepton sector together
with a small angle, $V_{cb}={\cal{O}}(m_s/m_b)$, in the quark sector from
the left diagonalization matrix. With some care, it is possible to couple
this picture, based on an asymmetric Yukawa coupling matrix, with hierarchical
neutrino masses and to extend it to the full three families. Several
variations of it, not always reducible to SU(5) language and anyhow mostly
employing abelian flavour symmetries, can be found in the literature.

At variance with this case, in this paper we explore the possibility
that $\theta_{23}={\cal{O}}(1)$ and $\Delta m^2_{23} \gg \Delta
m^2_{12}$ are explained in a context where the charged fermion mass
matrices, $m^u$, $m^d$, $m^e$, as the neutrino Dirac mass matrix
$m_{LR}$, do not have off-diagonal elements in the flavour basis which
are significantly asymmetric. More specifically, we look for an
extension to the neutrinos of the analysis of the charged fermion mass
matrices based on a U(2) flavour symmetry~\cite{chargedU2}.  In
Section \ref{chargedfermions} we summarize for easy of the reader the
main features of the U(2) analysis of charged fermion masses. In
Section \ref{neutrinos} we give conditions for incorporating the
relations $\theta_{23} = {\cal{O}}(1)$ and $\Delta m^2_{23} \gg \Delta
m^2_{12}$ in an extension of U(2) to the neutrino sector and we derive
its quantitative consequences. A possible realization of this general
pattern is described in Section \ref{example}. Conclusions are
summarized in Section \ref{conclusions}.

\section{\label{chargedfermions}U(2) and charged fermions masses}
The three family multiplets of matter fields $\psi_i$, $i$=1,2,3,
transform under U(2) as a doublet and a trivial singlet: $\psi_i =
\psi_a \oplus \psi_3$, $a$=1,2.  We view each $\psi_i$ as a {\bf 16}
of SO(10), therefore including a right-handed neutrino. The flavon
fields which can couple to the matter bilinears in a U(2)-invariant
way, a triplet $S^{ab}$, a doublet $\phi^a$ and an antisymmetric
singlet $A^{ab}$, break hierarchically the flavour group as: 
\begin{equation}
\label{breaking}
{\rm U(2)}\stackrel{\vev S,\vev\phi}{-\!\!\!-\!\!\!\longrightarrow}{\rm U(1)}
\stackrel{\vev A}{-\!\!\!-\!\!\!\longrightarrow}\{e\}
\end{equation} 
where U(1) corresponds, in an appropriate basis, to the subgroup of
phase rotations of the lightest family and $\{e\}$ is the unity of
U(2). More precisely if, in units of a basic scale $M$, 
\begin{equation}
\label{ordersofmag}
\frac{\left\|\vev S\right\|}{M} \simeq
\frac{\left\|\vev\phi\right\|}{M} \simeq \mep
\gg \frac{\left\|\vev A\right\|}{M} = \mepp,
\end{equation}
where \ep, \epp\ are two small dimensionless parameters, it is
possible to show~\cite{alignment}, under general conditions, that
$\vev S$ and $\vev\phi$ are misaligned in U(2) space only by a
relative amount of order \epp, or in an appropriate basis and up to
order one prefactors: 
\begin{equation}
\label{alignment}
\begin{array}{cc} \phi \simeq \left(\begin{array}{c} \mep\mepp \\ 
\mep\end{array}\right) &
S \simeq \left(\begin{array}{cc} {\mepp}^2 & \mep\mepp \\ \mep\mepp & 
\mep\end{array}\right).
\end{array}
\end{equation}  
Allowing for a general U(2)-invariant Yukawa coupling to Higgs fields
containing weak doublets VEVs, this gives the following structure of
the Yukawa coupling matrices in flavour space: 
\begin{equation}
\label{U2yukawa}
\frac{\boldsymbol{\lambda}}{\lambda_{33}} = \left(\begin{array}{ccc}
{\mepp}^2 & \mepp & \mep\mepp \\ \mepp & \mep & \mep \\ \mep \mepp & \mep & 1
\end{array}\right).
\end{equation}
For every entry the corresponding size of the U(2) breaking parameter
is indicated.  It is only in the case of the 12 and 21 entries that
U(2) implies a specific relation, $\lambda_{12} = -\lambda_{21}$, up
to correction of relative order \ep.

Al least to account for $m_c/m_t \ll m_s/m_b \simeq m_{\mu}/m_{\tau}$
and $m_u/m_t \ll m_d/m_b \simeq m_{e}/m_{\tau}$ a vertical structure
has to be supplemented in (\ref{U2yukawa}). Taking advantage of the
antisymmetry in flavour space of the 12, 21 couplings as opposed to
the symmetry of the 11, 22 elements, it is possible to further
suppress every entry of the entire 12-block of the $\lambda^u$ matrix
by an SU(5)-breaking parameter $\rho$~\cite{chargedU2}.  As an
example the flavon $A^{ab}$ may be an SO(10) singlet or a {\bf 45} of
SO(10) with a SU(5) symmetric VEV, whereas $S^{ab}$ can be a {\bf 45}
of SO(10) with a VEV in the B - L direction.  As long as there is no
other SU(5) breaking VEV, the 12-block of the $\lambda^u$ matrix
vanishes, since $u$ and $u^c$ belong to the same SU(5) multiplets,
unlike the case for $d$, $d^c$ or $e$, $e^c$ or $\nu_L$, $\nu_R$. At
the same time, up to SU(5) breaking corrections,
$\left|\lambda^d_{21}\right| = \left|\lambda^d_{12}\right| =
\left|\lambda^e_{21}\right| = \left|\lambda^e_{12}\right|$ and
$3\left|\lambda^d_{22}\right| = \left|\lambda^e_{22}\right|$, since
$d$, $e^c$ and $d^c$, $e$ live in the same SU(5) multiplets, whereas
no special relation is implied for the Dirac neutrino mass matrix
$\lambda^{LR}$, although also non vanishing in the SU(5) limit. 

In summary, we are led to the following dependence of the mass matrices on the
U(2) and SU(5) symmetry breaking parameters (all normalized to the scale M)
\ep, \epp\ and $\rho$ respectively:
\begin{equation}
\label{downyukawa}
\left(\frac{\boldsymbol{\lambda}}{\lambda_{33}}\right)_{d,e,LR} \simeq
\left(\begin{array}{ccc}
{\mepp}^2 & \mepp & \mep\mepp \\ \mepp & \mep & \mep \\ \mep\mepp & \mep & 1
\end{array}\right),
\end{equation}
\begin{equation}
\label{upyukawa}
\left(\frac{\boldsymbol{\lambda}}{\lambda_{33}}\right)_{u} \simeq
\left(\begin{array}{ccc}
{\mepp}^2\rho & \mepp\rho & \mep\mepp \\ \mepp\rho & \mep\rho & \mep \\ 
\mep\mepp & \mep & 1
\end{array}\right),
\end{equation}
with the particular relations: 
\begin{equation}
\label{12relation}
\lambda^{u,d,e,LR}_{12} = -\lambda^{u,d,e,LR}_{21},
\end{equation}
\begin{equation}
\label{22relation}
\left|\lambda^{e}_{22}\right| = 3\left|\lambda^{d}_{22}\right|,\qquad
\left|\lambda^{e}_{12}\right| = \left|\lambda^{d}_{12}\right|,\qquad
\left|\lambda^{e}_{33}\right| = \left|\lambda^{d}_{33}\right|.
\end{equation}
Allowing for prefactors of order unity in (\ref{downyukawa}) and
(\ref{upyukawa}) consistent with (\ref{12relation},\ref{22relation}),
all presently known properties of quarks and charged leptons are well
reproduced by these mass matrices with $\mep \simeq \rho \simeq 2
\times 10^{-2}$ and $\mepp \simeq 4 \times
10^{-3}$~\cite{chargedU2}. Eqs.\ (\ref{22relation}) give, in
particular, the well known Georgi-Jarlskog relations among fermion
masses~\cite{GJ}.  As shown elsewhere~\cite{precisetest}, the
relations~\cite{fritzsch}:
\begin{equation}
\label{test}
\displaystyle{\left|\frac{V_{ub}}{V_{cb}}\right| =
\sqrt{\frac{m_u}{m_c}}},\qquad
\displaystyle{\left|\frac{V_{td}}{V_{ts}}\right| = \sqrt{\frac{m_d}{m_s}}}, 
\end{equation}
implied by (\ref{downyukawa}--\ref{12relation}), valid up to
corrections of relative order \ep, represent a test in qualitative
agreement with present data, whose significance should improve
considerably in the near future.

\section{\label{neutrinos} Extension to neutrinos: general considerations }
The extension to neutrinos requires knowing the symmetry properties of
the right-handed neutrino mass matrix, $m_{RR}$, entering the see-saw
formula for the light neutrinos $m_{LL} =
m_{LR}m_{RR}^{-1}m_{LR}^{t}$. In general $m_{RR}$ arises from
$\overline{{\bf 126}}$ representations of SO(10), fundamental or
effective, also transforming under U(2) as singlets, $\Omega$,
doublets, $\Omega^a$, or triplets $\Omega^{ab}$. The U(2)
antisymmetric singlet does not couple to a neutrino bilinear.

Which structure of $m_{RR}$ could give a large $\theta_{23}$ angle and
the neutrino mass hierarchy? Note that $\theta_{23} = {\cal{O}}(1)$
should come from the diagonalization of $m_{LL}$ in view of the form
(\ref{downyukawa}) of $\lambda^e$. Note also that, if all the
$\Omega$'s were present with the maximal strength consistent with
U(2)-breaking, \ie\ $\|\vev\Omega\|/M = {\cal{O}}(1)$,
$\|\vev{\Omega^a}\|/M \simeq \left\|\vev{\Omega^{ab}}\right\|/M =
{\cal{O}}(\mep)$, the resulting $m_{LL}$ would not have any of the
desired properties.

In view of this, based on a classification of the possible forms of
$m_{LL}$ giving $\theta_{23} = {\cal{O}}(1)$ and $\Delta m^2_{23} \gg
\Delta m^2_{12}$~\cite{textures}, we consider the following ansatz for
the $\Omega$'s: i) The singlet $\Omega$ should be absent or
sufficiently suppressed; ii) $\|\vev{\Omega^a}\| \simeq
\left\|\vev{\Omega^{ab}}\right\| \simeq \mep M$ and, in the U(2) basis
where $\Omega^1 = 0$,
\begin{equation}
\label{omegaalign}
\frac{\Omega^{ab}}{M} \simeq \left(\begin{array}{cc}
0 & \mep\mepp \\ \mep\mepp & \mep 
\end{array}\right).
\end{equation}     
Again $\Omega^{11}=0$ means that $\Omega^{11}$ is sufficiently
suppressed.  As we shall see in the next Section, all this can be
explicitly implemented in at least one concrete example.

We argue that the resulting $m_{RR}$ gives the desired properties of
$m_{LL}$ under a suitable condition. It is, in the same flavour basis
as~(\ref{downyukawa}),
\begin{equation}
\label{mrr}
m_{RR}=M\left(\begin{array}{ccc} 0 & \mep\mepp & 0 \\
\mep\mepp & \mep & \mep \\ 0 & \mep & 0\end{array}\right),
\end{equation}
which has, in this approximation, a massless eigenstate:
\begin{equation}
\label{light}
N_l \simeq N_1 + {\cal{O}}(\mepp)N_3
\end{equation} 
and two heavy states, $N_h^1$ and $N_h^2$, of similar masses
$M_{h_1}$, $M_{h_2}$, predominantly composed of
$N_2$ and $N_3$ with comparable coefficients. If expressed in the
basis of these right-handed neutrinos mass eigenstates, $N_R =
\left(N_l, N_h^1,N_h^2\right)$, the Dirac mass matrix $m_{LR}$ of
(\ref{downyukawa}) acquires the form:
\begin{equation}
\label{mLRnewbase}
\frac{m'_{LR}}{m_{LR}^{'33}} \simeq \left(\begin{array}{ccc}
{\mepp}^2 & \mepp & \mepp \\ \mepp & \mep & \mep \\ \mepp & 1 & 1
\end{array}\right).
\end{equation}
Notice that we have achieved a lighter right-handed neutrino $N_l$,
massless in the exact limit of (\ref{mrr}), predominantly coupled to
$\nu_{\mu}$ and $\nu_{\tau}$ with comparable strength.  This is the
key for having a large $\theta_{23}$ angle and, at the same time,
hierarchical left-handed neutrinos~\cite{light}. If $N_l$ is light
enough, $M_l \ll M_h$, the dominant terms from $N_R$ exchanges in the
mass Lagrangian of the left-handed neutrinos are:
\begin{equation}
\label{dominantterm}
{\cal{L}}_{m_{LL}} \simeq \frac{v^2}{M_l}{\mepp}^2\left(\nu_{\mu}+\nu_{\tau}
+ \mepp\nu_e\right)^2 + \frac{v^2}{M_h}\nu_{\tau}^2 
\end{equation}
where we have set $m_{LR}^{33} \simeq v$, a typical SU(2) $\times$
U(1) breaking vacuum expectation value, and all terms are meant to
have a numerical coefficient of order unity. Therefore, if
\begin{equation}
\label{condition}
\frac{M_l}{M_h} \ll {\mepp}^2,
\end{equation}
the left-handed neutrino masses are hierarchical:
\begin{equation}
\label{leftneutrinomasses}
m_3 \simeq \frac{v^2{\mepp}^2}{M_l} \gg m_2 \simeq \frac{v^2}{M_h}
\end{equation}
and $m_{LL}$ is diagonalized by an order 1 rotation in the
$\nu_{\mu}/\nu_{\tau}$ - $2/3$ sector, up to further rotations with
angles of order \epp.

The fact that ${\cal{L}}_{m_{LL}}$ in~(\ref{dominantterm}) is
diagonalized, to a good approximation, by an order 1 rotation in the
$\nu_{\mu}/\nu_{\tau}$ sector:
\begin{equation}
\label{neutrrotation}
V^{\nu} \cong \left(\begin{array}{ccc}
1 & {\cal{O}}(\mepp) & {\cal{O}}(\mepp) \\ {\cal{O}}(\mepp) & \cos\bar\theta &
 \sin\bar\theta \\ {\cal{O}}(\mepp) & -\sin\bar\theta & \cos\bar\theta
\end{array}\right),
\end{equation} 
together with the explicit form of $\lambda^e$ in
eqs.~(\ref{downyukawa}) and (\ref{12relation}), has an important
implication. The mixing matrix in the leptonic charged weak current
$\overline{\ell}\gamma_{\mu}V^{\ell}\nu$ is in fact given by:
\begin{equation}
\label{totalrotation}
V^{\ell}=(V^E)^{\dagger}V^{\nu},
\end{equation}
where $V^E$ is the left unitary rotation needed to diagonalize
$\lambda^e$. In view of (\ref{downyukawa},\ref{12relation}) it is:
\begin{equation}
\label{chargedrotation}
V^E \cong
\left(\begin{array}{ccc}
1 & 0 & 0 \\ 0 & 1 & {\cal{O}}(\mep) \\ 0 & {\cal{O}}(\mep) & 1 
\end{array}\right)
\left(\begin{array}{ccc}
\cos\theta_E & \sin\theta_E & 0 \\ -\sin\theta_E & \cos\theta_E & 0 \\
0 & 0 & 1
\end{array}\right),
\end{equation}
where $\tan\theta_E = \sqrt{(m_e/m_{\mu})}$. Therefore we obtain a
mixing matrix:
\begin{equation}
\label{explicitrotation}
V^{\ell} \cong \left(\begin{array}{ccc}
\cos\theta_E & \sin\theta_E\cos\theta & \sin\theta_E\sin\theta \\
-\sin\theta_E & \cos\theta_E\cos\theta & \cos\theta_E\sin\theta \\
0 & -\sin\theta & \cos\theta
\end{array}\right)
\end{equation}   
or, adopting the common notation for $V^{\ell}$ in terms
of 2 $\times$ 2 rotations, 
\begin{equation}
V^{\ell} = 
R_{23}(\theta_{23}) R_{13}(\theta_{13})R_{12}(\theta_{12}),
\end{equation}
one has
\begin{equation}
\label{angles}
\theta_{12} = {\displaystyle \sqrt\frac{m_e}{m_{\mu}}
\cos\theta_{23}}, \qquad {\displaystyle \theta_{13} 
= \sqrt\frac{m_e}{m_{\mu}}\sin\theta_{23}}.
\end{equation}
These relations between $\theta_{12}$, $\theta_{13}$ and $\theta_{23}$,
valid up to correction of relative order \ep, are the analogue in the
lepton sector of eqs.\ (\ref{test}). For appropriate values of the
neutrino masses, as discussed below, they are consistent with a
neutrino oscillation interpretation of the atmospheric and solar
neutrino anomalies, as known today. In particular, from $\sin^22\theta_{23} >
0.9$ we obtain
\begin{equation}
\label{sunprediction}
\sin^22\theta_{12} = (6 \div 13)\times 10^{-3},
\end{equation}     
well compatible with the small angle MSW interpretation of the solar
neutrino data\footnote{The second of (\ref{angles}) implies
$\theta_{13} = (2 \div 3)^\circ$ compatible with the CHOOZ limit,
$\theta_{13} < 13^\circ$ if $\Delta m^2_{\rm atm} > 2 \times 10^{-3}\:
{\rm eV^2}$~\cite{CHOOZ}, and with the somewhat less stringent limits
from Super-Kamiokande: $\theta_{13} \lesssim
20^\circ$~\cite{theta13}.}~\cite{sun}.

\section{\label{example} An explicit example}
In this Section we briefly discuss an explicit realization of the
picture described previously. Other than showing a concrete example,
there are two related reasons for doing this. The zeros in (\ref{mrr})
will in general be replaced by small but non-vanishing entries. In turn
these entries determine if condition (\ref{condition}) is satisfied
and, at the same time, fix the order of magnitude of the neutrino
masses via (\ref{leftneutrinomasses}). Since condition
(\ref{condition}) requires:
\begin{equation}
\label{detcondition}
\det m_{RR} = M_{h}^1M_{h}^2M_{l} \ll \mep^3{\mepp}^2M^3,
\end{equation} 
barring cancellations among terms of the same order of magnitude, it must be:
\begin{equation}
\label{entrycondition}
m_{RR}^{33} \ll M\mep,\qquad m_{RR}^{11} \ll M\mep{\mepp}^2,\qquad
m_{RR}^{13} \ll  
M\mep\mepp.
\end{equation} 

A possible concrete realization is obtained by assuming that
$\Omega^a$ and $\Omega^{ab}$ are both fundamental fields, transforming
as ${\bf \overline{126}}$ of SO(10), to be added to the flavons
$\phi^a$, $S^{ab}$ and $A^{ab}$ introduced in Section
\ref{chargedfermions}, that couple to charged fermions bilinears.  In
close analogy with the discussion made in ref.~\cite{alignment}, by
considering an SO(10) $\times$ U(2) invariant potential which also
includes all flavon fields with opposite SO(10) $\times$ U(2)
transformation properties, denoted by a bar, it is possible to show
that the appropriate SU(3) $\times$ SU(2) $\times$ U(1) invariant
components of $\Omega^a$, $\Omega^{ab}$, $\phi^a$ are aligned in U(2)
space as:
\begin{equation}
\label{126alignment}
\Omega^a \simeq \phi^a,\qquad \Omega^{ab} \simeq 
{\displaystyle \frac{1}{M\mep}\:(\Omega^a\phi^b+\Omega^b\phi^a}), 
\end{equation} 
to a sufficiently good approximation, so that (\ref{entrycondition})
are fulfilled\footnote{The same approximate alignment holds for the
barred fields.}.  More precisely in the U(2) basis where $\Omega^a = M
(0,\mep)$, eqs.~(\ref{alignment},\ref{omegaalign}) hold and the
dominant correction to (\ref{mrr}) arise from the couplings:
\begin{gather}
\frac{1}{M}\nu_{R3}(\Omega^a\bar{\phi}_a +
\Omega^{ab}\bar{S}_{ab})\nu_{R3}, \notag \\ 
\frac{1}{M}\nu_{Ra}(\Omega^{ab}\bar{\phi}_b +
\frac{1}{M}A^{ab}\Omega^c\bar{S}_{bc})\nu_{R3},\\ 
\frac{1}{M^2}\nu_{Ra}(S^{ab}\Omega^c\bar{\phi}_c +
A^{ac}\Omega^{bd}\bar{S}_{cd})\nu_{Rb}, \notag
\end{gather}
which give respectively:
\begin{equation}
\label{newcontribution}
\begin{array}{ccc}
m_{RR}^{33} \simeq M\mep^2 & m_{RR}^{13} \simeq M\mep^2\mepp &
m_{RR}^{11} \simeq M\mep^2{\mepp}^2
\end{array}.
\end{equation}
One has therefore:
\begin{equation}
\label{finalmrr}
m_{RR} \simeq M\mep\left(\begin{array}{ccc}
\mep{\mepp}^2 & \mepp & \mep\mepp \\ \mepp & 1 & 1 \\ \mep\mepp & 1 & \mep
\end{array}\right) 
\end{equation}
which gives $M_{h}^1 \simeq M_{h}^2 \simeq \mep M$ and $M_l \simeq
\mep^2{\mepp}^2 M$ or, from eqs.~(\ref{leftneutrinomasses}):
\begin{equation}
\label{leftmassvalues}
{\displaystyle m_3 \simeq \frac{v^2}{M\mep^2}},\qquad {\displaystyle 
\frac{m_2}{m_3} \simeq \mep}.
\end{equation} 
Taking $v=250\: {\rm GeV}$, $M=M_{\rm Planck}$ and $\mep = 2 \times
10^{-2}$ as required to describe the quark
parameters~\cite{chargedU2}, we obtain
\begin{equation}
\label{squaredmasses}
\begin{array}{c}
\Delta m^2_{23} \simeq m_3^2 = {\cal{O}}(10^{-2}\: {\rm eV}^2) \\ \\
\Delta m^2_{12} \simeq m_2^2 = {\cal{O}}(10^{-5} \div 10^{-6}\: {\rm eV}^2) 
\end{array}
\end{equation}
which, together with (\ref{angles}), can give a consistent
description of atmospheric and solar neutrino data so far. Note that
the lightest right-handed neutrino has a mass of order $10^{10}\:{\rm
GeV}$. 

Before closing this Section, we would like to comment on the
possibility of accommodating in a U(2) model alternative patterns of
neutrino masses and mixings than the one described so far, always
accounting in a parametric way for $\theta_{23} = {\cal{O}}(1)$ and
$\Delta m^2_{23} \gg \Delta m^2_{12}$\footnote{Alternative U(2) models
not trying to incorporate parametrically $\theta_{23}={\cal O}(1)$ and
$\Delta m^2_{23} \gg \Delta m^2_{12}$ in a 3 neutrino scheme can be
found in refs.~\cite{U2neutrinos}.}. This is a difficult question to
answer in general. We have been able to find, however, an alternative
example always based on the dominance of a light right-handed
neutrino, coupled with comparable strength to $\nu_{\mu}$ and
$\nu_{\tau}$. In this model we have, besides the $\phi$, $S$, $A$
flavons, a fundamental $\overline{\bf 126}$, $\Omega_A^{ab}$, which is
an antisymmetric tensor under U(2) and has a vacuum expectation value
$\left|\Omega_A\right| \simeq \mepp M$. In this case the dominant
operators contributing to the different entries of $m_{RR}$ are as
follows:
\begin{equation}
\label{antisymmMrr}
m_{RR} \simeq \Omega^{12}_A \left(\begin{array}{ccc}
\bar{S}_{22}A^{12} & \bar{S}_{22}S^{22} + \bar \phi_2\phi^2 & \bar{\phi}_2 \\
\bar{S}_{22}S^{22} + \bar \phi_2\phi^2 & \bar{A}_{12}S^{22} & \bar{\phi}_1 + 
\bar A_{12}\phi^2\\
\bar{\phi}_2 & \bar{\phi}_1 + \bar A_{12}\phi^2 & \bar{A}_{12}
\end{array}\right)
\end{equation} 
so that:
\begin{equation}
\label{antisymmMrrNum}
m_{RR} \simeq M\mep\mepp\left(\begin{array}{ccc}
\mepp & \mep & 1 \\ \mep & \mepp & \mepp \\ 1 & \mepp & \mepp/\mep
\end{array}\right).
\end{equation}
When inserted in $m_{LL} = m_{LR}m_{RR}^{-1}m_{LR}^t$ together with
(\ref{downyukawa}) this gives:
\begin{equation}
\label{antisymmpred1}
\theta_{23} \simeq 1,\qquad {\displaystyle \theta_{12} \simeq
\theta_{13} \simeq \frac{\mepp}{\mep}}
\end{equation}
and
\begin{equation}
\label{antisymmpred2}
{\displaystyle m_3 \simeq \frac{v^2\mep}{M{\mepp}^2}},\qquad {\displaystyle 
\frac{m_2}{m_3} \simeq \left(\frac{\mepp}{\mep}\right)^2}.
\end{equation}
Even though we loose the exact predictions (\ref{angles}), a
qualitative description of the data may be possible also in this case,
with appropriate ${\cal{O}}$(1) prefactors.

\section{\label{conclusions}Conclusions}
In this paper we have attempted a unified description of quark and
lepton masses and mixings, based on a U(2) flavour symmetry. More
precisely, we have extended a previously developed description of the
flavour parameters of the charged fermions to include two main
features of the neutrino sector seemingly implied by recent
experimental findings: a large mixing angle between $\nu_{\mu}$ and
$\nu_{\tau}$ and a large hierarchy in the neutrino squared mass
differences, relevant to the oscillation phenomena. We have shown
that this is possible using an interplay between the flavour U(2)
symmetry and the vertical SO(10) symmetry. All the qualitative
features of the spectra and mixings of quarks and leptons can be
accounted for by two small parameters $\rho \simeq \mep$ and \epp\
expressing respectively the breaking scales of SO(10) $\times$ U(2)
to ${\rm SU}_{3,2,1} \times {\rm U(1)}$ and to ${\rm SU}_{3,2,1}$
relative to a basic scale $M$, close to the Planck mass.

From a phenomenological point of view, the neatest quantitative
predictions in the neutrino sector are given in eqs.~(\ref{angles}) or,
independently from the chosen parametrization of the mixing matrix in
the lepton sector, 
\begin{equation}
\label{finalprediction} 
{\displaystyle \left|V_{\mu 1}^\ell\right| =
\left|\frac{V_{e3}^{\ell}}{V_{\mu3}^{\ell}} \right| =
\left|\frac{V_{e2}^{\ell}}{V_{\tau3}^{\ell}}\right|
= \sqrt{\frac{m_e}{m_{\mu}}}},  
\end{equation} 
with a negligibly small CP-phase.
These relations, similarly to the analogous relations in the quark
sector, eqs.~(\ref{test}), should be valid to a good approximation and
should allow a quantitative test of the overall picture. In a specific
realization, we find also the order of magnitude of the neutrino
squared mass differences given in eqs.~(\ref{squaredmasses}). All of
these relations are compatible with the present experimental
information both in the quark sector and in the lepton sector.  Precise
comparisons should be possible in a not too distant future.

\end{document}